# Estimation of fatigue life of TiN coatings using cyclic micro-impact testing

Abdalrhaman Koko[1a], Elsiddig Elmukashfi[b], Tony Fry[a], Mark Gee[a], and Hannah Zhang[a]

[a] National Physical Laboratory, Hampton Road, Teddington TW11 0LW, UK

[b] School of Engineering, University of Leicester, Leicester LE1 7RH, UK

## Abstract

We studied the behaviour of a thin titanium nitride (TiN) coating (1.5 μm thick) on a tool steel substrate under dynamic and cyclic impacts through an approach combining experimental testing and computational modelling. Dynamic impact testing was used to investigate the load-dependent dynamic hardness and assess the energy-dissipation capabilities of the coating system. In cyclic impact tests, the coating-substrate system experienced permanent plastic deformation in each cycle, ultimately leading to coating failure. Chemical analysis identified an interlayer between the coating and the substrate that influenced the coating response, while cross-sectional analysis revealed the extent of coating damage due to cycling and impact load. A three-dimensional map was constructed, connecting the acceleration load, sensed depth, and cycles to the coating failure, and an empirical equation was used to characterise the relationship between the depth and cycles to failure. The computational model examined the traction component distribution during loading and unloading, focusing on normal and shear tractions. These findings suggested the potential significance of normal traction in the interfacial fatigue failure due to impact.

**Keywords:** Impact testing; fatigue; titanium nitride; coating; S-N curve; dynamic hardness

---

[1] Corresponding author. E-mail address: abdo.koko@npl.co.uk

# 1. Introduction

Coatings are generally designed to extend the performance and lifetime of tools and components or impart additional properties or functionality. The durability of the coating is critical to its usefulness; therefore, wear resistance and interfacial strength are essential parameters to measure and characterise. Fatigue failure is a common failure mechanism in materials subjected to repeated cyclic loading, leading to accumulative damage within the material and, ultimately, catastrophic failure. Where coatings are used on machining tools, fatigue wear or fatigue contact failure owing to cyclic stress can significantly reduce the tool's life, leading to increased maintenance costs and downtime [1], [2], [3]. Thus, estimating the fatigue life is an important aspect of evaluating the performance and durability of coatings.

Several studies have investigated the fatigue behaviour and mechanical properties of TiN coatings. In terms of contact fatigue performance, Xu *et al*. [4] investigated the effects of TiN coatings on gears. They found that TiN-coated gears exhibited an improved contact fatigue life and higher contact fatigue strength compared to uncoated gears. Costa *et al*. [5] studied the uniaxial fatigue behaviour of a TiN-coated Ti-6Al-4V alloy. They found that using a TiN coating improved the fatigue limit of the alloy by 60 MPa without significant changes in the yield and tensile strength. On the other hand, Zhang et al. [6] investigated the fatigue and mechanical behaviour of a Ti-6Al-4V alloy with TiN and CrN coatings. They observed that the fatigue strength of the coated samples was lower than that of the uncoated substrate in both low- and high-cycling fatigue regimes. The reduction in the fatigue strength was more significant for the CrN coating owing to its inferior plastic deformation capacity. However, the coated samples exhibited higher ultimate and yield strengths than those of the uncoated substrate.

In terms of adhesion, Khlifi and Larbi [7] conducted scratch tests to evaluate the critical forces of adhesion for various physical vapour deposition (PVD) coatings, including TiN. They found that the TiN coating had the lowest critical adhesion force compared with the other coatings. Additionally, impact testing revealed that TiN coatings displayed circumferential and radial cracks during impact-cyclic loading. Uddin *et al*. [8] investigated the tribological and mechanical properties of TiN coatings on AISI 52100 bearing steels. They varied the substrate surface roughness and coating thickness and performed scratch and tribological tests. The results showed that samples with a specific surface roughness and coating thickness exhibited better adhesion, wear resistance, and a lower coefficient of friction for the TiN coating. Gopkalo and Rutkovskyy [9] studied the influence of PVD coatings, including TiN, on the tensile strength and low-cycle fatigue resistance of stainless steel and titanium alloys. They found that the PVD coatings formed hard, stronger layers with high residual compressive stresses, which increased their tensile strength and resistance to fatigue loading.

Recently, more attention has been paid to the fatigue behaviour of coatings and the progressive weakening of the initial coating-substrate interfacial strength, focusing on in situ testing [3]. However, there is still a need for a greater understanding of the relationship between coating fatigue strength and the coating-substrate interfacial strength. This is because evaluating the toughness of thin coatings, especially hard coatings, is still challenging due to the specimen dimensions, difficulty in accurately replicating the complex loading conditions experienced by materials in real-world applications, and lack of reliable test procedures [10].

Impact testing provides an efficient experimental method for evaluating the fatigue resistance of the coatings. It can be used to investigate the fatigue strength of a coating surface using different acceleration loads, thereby inducing higher localised stresses. A cyclic impact test (strain rate of $10^3 - 10^4$/s), typically using a spherical indenter, was developed as a practical experimental methodology to assess the *impact* fatigue strength of hard coatings exposed to different and repetitive mechanical impact loads [11]. A spherical indenter provides a more uniform stress distribution under the indenter tip compared to sharper indenter shapes [12], and the non-singular nature of the stress fields produced by spherical indenters, which makes the analysis and testing less complex and more suitable for the mathematical modelling of indentations of thin, functionally graded, or multi-layered coatings [13].

Micro-impact testing, compared to macro-impact testing, is suitable for a coating thickness ($t$) to spherical indenter radius ($R$) ratio ($t/R$) of 0.1, and can provide accurate depth sensing, time-to-failure, measurement of coating mechanical properties and adhesion, and overall provides a more localised assessment of impact resistance [11]. Thus, this study provides a systematic experimental approach for investigating the fatigue properties of thin TiN coatings on tool steel using repeated micro-impact testing, with a range of acceleration loads to simulate low-cycle fatigue conditions. We used experimental data to extract a load-cycle curve description for coating fatigue life, describe coating failure mechanics, and outline how this can be incorporated into future models to provide a single descriptor for coating failure.

## 2. Methodology

### 2.1. Coating procedure

Titanium Nitride (TiN) coating was applied to $2 \times 2 \times 0.5$ cm$^3$ samples using an Electron Beam Physical Vapor Deposition (EB-PVD) system by Wallwork®. The samples were cleaned in a heated alkaline ultrasonic bath for 16 mins, followed by deionised water rinse and hot nitrogen dry. The sample was further cleaned using an IPA wipe before fixturing and placing it into the EB-PVD coating chamber. The chamber was then evacuated to a vacuum level better than $10^{-5}$ mbar and heated to 400 °C for 90 minutes. Sputter cleaning was achieved by backfilling the chamber with high-purity argon to $1.5 \times 10^{-2}$ mbar for 10 minutes at 1 kV.

The coating was applied by first depositing pure Ti under 30 sccm of high-purity argon and a substrate bias of -50 V to achieve a thickness of less than 300 nm before introducing 80 sccm of high-purity nitrogen to deposit TiN. High-purity nitrogen and argon gases, controlled by mass flow controllers, were used as the reactive and working gases, respectively, and the electron beam evaporated the titanium target, and the titanium ions which then combined with the nitrogen to form TiN, which was deposited on the substrate.

The timing was adjusted to deposit a nominal 1.50 ± 0.05 µm thick layer of TiN (Figure 5b). The temperature was maintained at 400 °C for the duration of this process. Unlike other methods, such as ARC deposition, the EB-PVD process produces a smooth finish with a more homogeneous structure.

### 2.2. Impact testing

Pendulum-based micro-impact testing was performed using a NanoTest Xtreme instrument (Micro Materials Ltd.), fitted with a 4.37 µm radius diamond spherical indenter with the radius being measured by scanning electron microscopy (SEM). The spherical part of the indenter had a depth of approximately 1.8 µm (from the tip) until it became a 90° cone.

In the dynamic hardness test (Figure 1a), the impact was produced by using a solenoid to raise the pendulum armature to its raised position; the solenoid was then turned off, releasing it and allowing the pendulum to accelerate towards the specimen surface with a measured acceleration to induce a specific load called the *acceleration load*. The indenter displacement along the acceleration arc and during the impact was recorded with a 10 kHz data acquisition frequency, allowing for the indenter velocity and impact instantaneous strain rate to be measured and calculated, respectively [14]. A range of incident velocities spanning nearly two orders of magnitude was achieved by varying the acceleration and the spacing between the solenoid and pendulum [15]. In addition, the maximum material penetration, residual deformation, and the ability to plastically deform the coating were calculated from the displacement and time data [16]. The dynamic hardness tests were repeated five times.

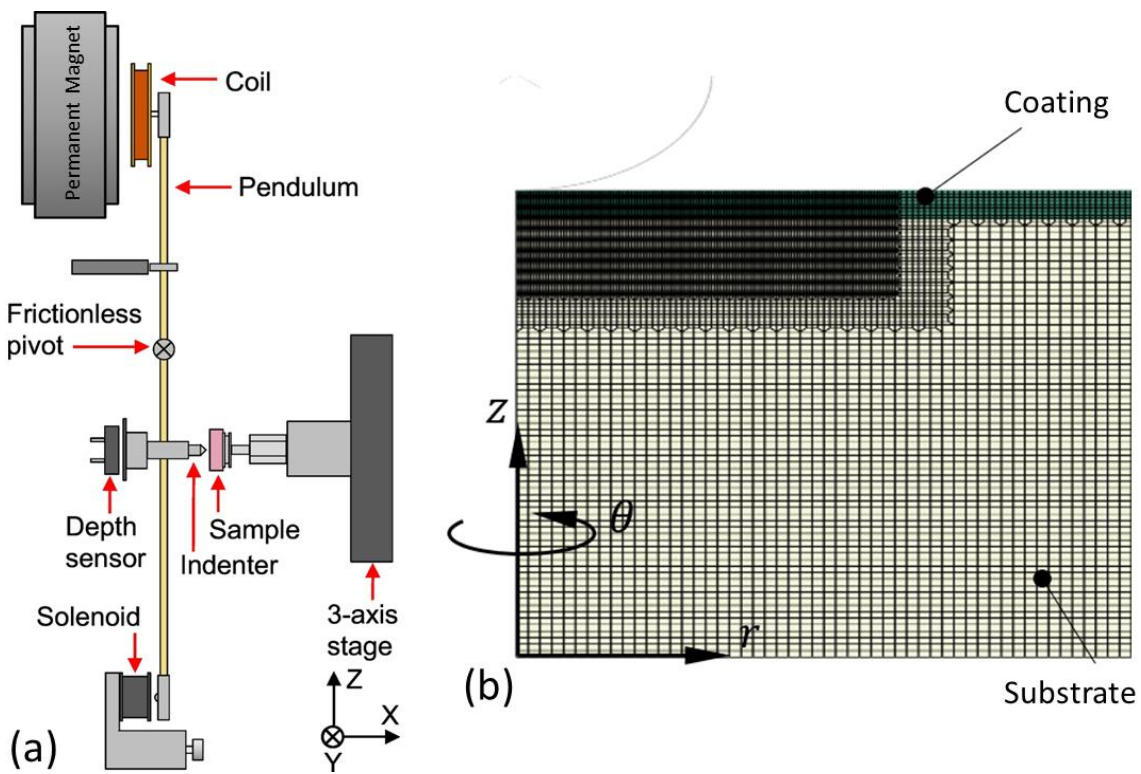

Figure 1: (a) Schematic of the Pendulum-based micro-impact testing, culled from Rueda-Ruiz et al. [15]. (b) Finite element model of indentation experiment.

Dynamic impact testing describes energy loss, indentation depth recovery, and dynamic hardness over time [16]. Impact testing has been demonstrated as an effective approach for measuring the dynamic hardness of materials. The dynamic hardness ($H_D$) of the coating-substrate system was calculated as the ratio of the plastic work done by the impact ($W_{\text{plastic}}$) to the impact crater volume ($V_{\text{crater}}$) [17].

$$H_D = \frac{W_{\text{plastic}}}{V_{\text{crater}}} = \frac{1}{2} m \frac{(v_{in}^2 - v_{out}^2)}{V_{\text{crater}}} \qquad 1$$

where $v_{in}$ and $v_{out}$ are the velocities at which the indenter approaches and recedes from the surface after impact, respectively, calculated from the change in depth with time [15]. $m$ is the effective system mass, which was 240 g for the current setup. For current set up, the surface between indenter and surface is 27 µm.

Repetitive cyclic impact tests were performed to investigate the low-cycle fatigue behaviour of the coating system under a high strain rate. Repeated impacts were conducted using the same acceleration parameters as those used for single-impact tests, repeating impacts at the exact location up to 7,500 times, with a time interval of 4 s to allow the ringing in the pendulum to dissipate. This ringing, due to the bouncing of the spherical indent hitting the sample, should not be confused with the free oscillation of the pendulum; see Figure 1a and [18] for more details. After moving 50 µm from the previous impact site to ensure no interaction between the indents, the acceleration load was increased following the Fibonacci sequence ($L_n = L_{n-1} + L_{n-2}$) starting from 5 mN, and the test was repeated up to 100 mN.

## 2.3. Postmortem SEM Characterisation

The diameter of the impact crater was measured from high-resolution SEM images acquired using a Zeiss Auriga dual-beam SEM system at 3 kV and 30 µm aperture size with a pixel size of 6 nm using ImageJ software [19]. The measurement was repeated five times at different circumferential positions around the impact crater.

Selected impact craters were then sectioned by milling trenches orthogonally to the surface using a Zeiss Auriga dual-beam SEM system fitted with a Schottky field emission Gemini electron column and an Orsay Physics Ga+ ion field ion beam. The focussed Ion Beam (FIB) milling was done using 4 nA/ 30 kV conditions before a fine current of 240 pA/ 30 kV produced a smooth surface with limited FIB curtaining. Finally, the surface of the FIB trench wall was observed using mixed in-lens and secondary electron (SE) imaging (75% and 25%, respectively) to enhance feature detection. The selected impact craters were chemically analysed using Oxford Instruments' Ultim energy-dispersive X-ray spectroscopy (EDX) at 10 kV, 60 µm aperture size, and 5 mm working distance.

## 2.4. Computational model

Finite element analysis (FEA) of the indentation experiment was performed using ABAQUS® software (version 6.6). An axisymmetric finite-element model was used (Figure 1b). The model was divided into two regions: the TiN coating and the ASP-23 tool steel substrate. The width and height of the half-space were taken as five times the indenter radius, which is essential for avoiding boundary effects. A uniformly refined element region was created beneath the indenter in the coating and substrate to control element length. 4-node bilinear axisymmetric elements (CAX4H) were used to discretise the axisymmetric conditions. It should be mentioned that a hybrid formulation was adopted to improve the convergence at large strains, where incompressible plastic deformation dominates. The mesh had 14,400 elements with a controlled element length of 0.09 μm in the refined region, necessary to obtain an optimal solution, as confirmed by a convergence study that varied the element size over the ranges $l_e \in [0.05, 0.09, 0.20]\ \mu m$.

A rigid ball indenter was used in the model, and hard contact was assumed to be in the normal direction. A friction coefficient of 0.65 was used to allow for sliding [20]. The constitutive behaviour of the coating and substrate materials was assumed to be described by an isotropic von Mises plasticity model. The hardening function was considered as a power law, that is, $\sigma_y = \sigma_0 \left(1 + \varepsilon_e^{pl}/\varepsilon_0 \right)^{1/n}$ where $\sigma_0$ is the initial yield strength, $\varepsilon_e^{pl}$ is the equivalent plastic strain, $\varepsilon_0 = \sigma_0/E$, $E$ is Young's modulus, and $n$ is the hardening exponent. The coating material parameters were measured experimentally. The coating hardness was 27 ± 3 GPa, the reduced modulus was 319 ± 20 GPa, and the Young's modulus was 407 ± 19 GPa, as measured by Berkovich nanoindentation under the first condition mode (maximum indentation depth of 300 nm or maximum load of 30 mN) in accordance with ISO 14577 using Oliver-Pharr analysis. The indenter tip for the Berkovich static hardness and spherical dynamic hardness tests were shape-calibrated using a fused silica reference sample, and the instrument compliance was calibrated using a tungsten reference sample.

The hardening exponent ($n$) and the initial yield strength ($\sigma_0$) were obtained from Guang-yu He et al. [21], who used a FEM-reverse algorithm to quantify the elastoplastic properties of a 30 μm TiN coating, where $\sigma_0 = 18.25$ GPa, $\nu = 0.25$, and $n = 0.22$. The 30 μm thickness combined with nanoindentation provides a suitable thickness to minimise the possibility of delamination between the coating and substrate during nanoindentation compared to the 1.5 μm, which in turn will give an accurate representation of the coating that is separate from the substrate. In addition, typical parameters for tool steel were used, where $E = 230$ GPa, $\nu = 0.3$, $\sigma_0 = 1.84$ GPa, and $n = 0.47$ [22]. Note that in the model, the properties of the coating and steel were added separately to understand the coating-substrate system or 'coating system' response to impact that was probed experimentally.

Mechanical loading was applied as a prescribed force on the indenter in the $z$-direction. The force was ramped monotonically from zero to the maximum force, and then unloaded to zero. Rate-sensitivity effects may occur because of the higher loading rates in the experiment, but the model was limited to only considering the quasi-static case and focused on understanding the forces that drive the fatigue failure.

## 3. Results and discussion

The microstructure of a thin 1.5 μm Titanium Nitride (TiN) coating applied using Electron Beam Physical Vapor Deposition (EB-PVD) is characterised by small columnar grains. The size of these grains was in the range between 0.15 μm and 1.5 μm. The EB-PVD method is known for producing coatings with a columnar microstructure, which is crucial for strain tolerance and the lifetime of the coating. This columnar microstructure provides excellent strain tolerance and thermal shock resistance. In addition, using the EB-PVD produced a smooth golden finish with a more homogeneous structure.

### 3.1. Dynamic impact test

The pendulum was restrained using a solenoid during the impact loading, while an accelerating load was applied at the force coil. When the impact solenoid is turned off, the pendulum swings forward under an applied accelerating load across an acceleration distance of 27 μm (Figure 1a). The drag effect of the single damping plate can be observed in the plateau observed in the incident velocity as a function of the accelerating load (Figure 2a), which was measured and considered during further analysis. Given that the displacement of the indenter along the acceleration distance and during impact was measured using a high data logging frequency; the velocity before ($v_{in}$) and after ($v_{out}$) the first contact of the indenter were inferred (Table 1) via polynomial fits of the discrete differential of the approach portion of the displacement data as a function of the impact load, that is before the indenter contacted the sample at a displacement of ≥ 0 nm.

The depth-time profile (Figure 2a) indicates that the indenter bounced off the surface of the specimen multiple times (i.e., the ringing effect) until it reached a resting position against the specimen surface under an acceleration load. However, it was previously concluded that plastic deformation only occurred during the initial impact. In contrast, the secondary indenter impacts (i.e., bounces owing to the ringing effect) did not produce plastic deformation [23].

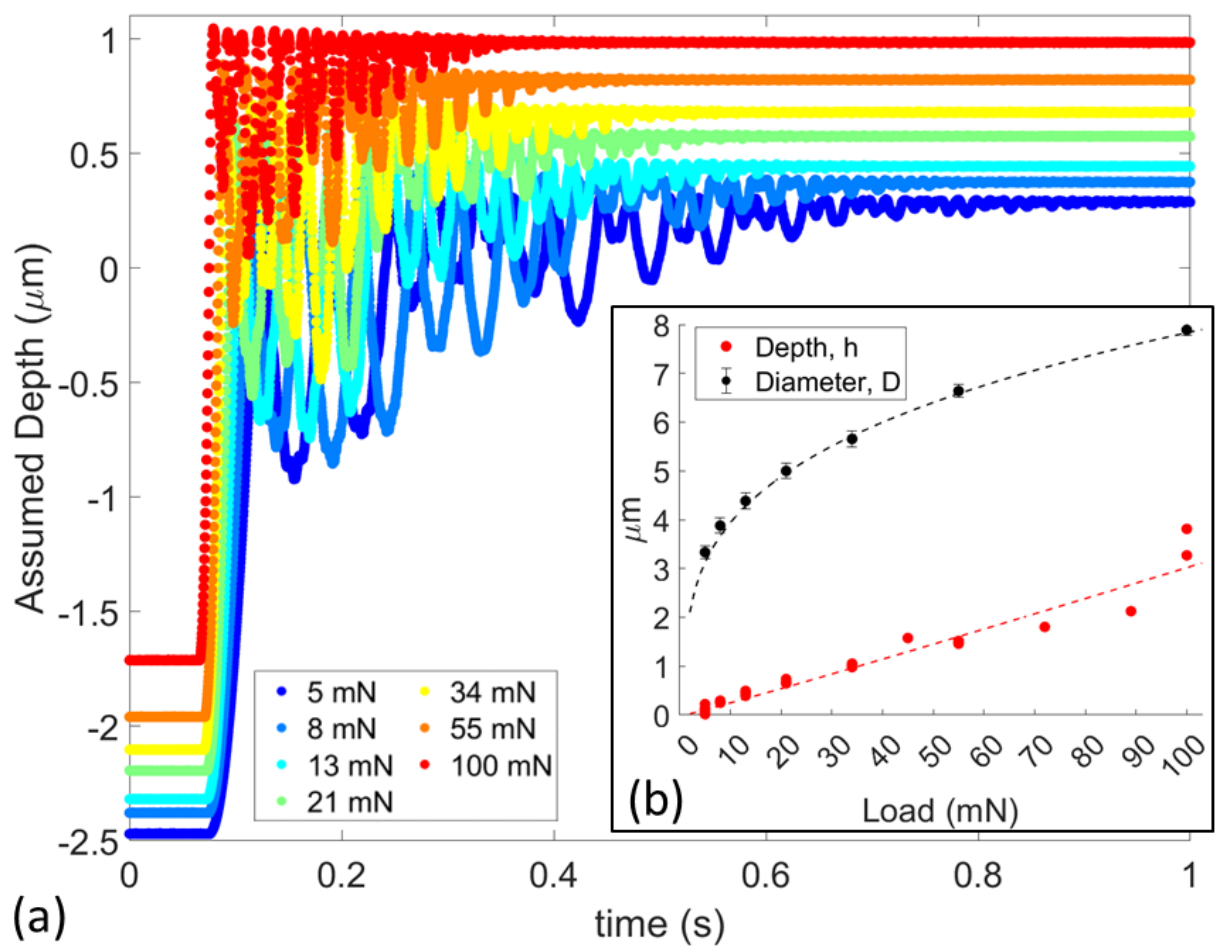

Figure 2: a) Assumed impact depth measured during the dynamic impact test for individual tests at seven acceleration loads. Plus and minus depth values describe the pendulum swinging motion, with zero being the crater's inferred depth. b) The measured depth ($h$) from the impact testing plotted against the impact crater's diameter was measured from SEM images of multiple tests at the seven loads.

Overall, this measurement approach depends on the calculation of the surface before impact, as detailed in [17] and [15]. This calculation changes when the setup is adjusted or recalibrated. This change may be due to influences from the solenoid-powered impact methodology, the application of a small load to locate the surface, or a change in the surface-to-indenter distance owing to the surface roughness. The calculation of the crater depth is also complicated by the ringing effect (shown in Figure 2a) when a spherical indenter hits the surface, and the maximum depth decreases during unloading owing to the elastic component of the induced deformation. In addition, the load could not be instrumented during the test, which could increase the accuracy of the depth sensing.

This issue can be circumvented using a partially filled spherical tank analogy, where once the level of water depth or diameter is known, the other unknown can be calculated using equation 2, where $R$ is the indenter radius, $D$ is the impact crater diameter measured using SEM images and ImageJ® (Figure 5a), and $h_{th}$ is the theoretical depth of the crater.

$$h_{th}(D) = \underbrace{R - \sqrt{R^2 - \frac{D^2}{4}}}_{\text{Spherical Tank}} = \underbrace{\frac{D^2 + R^2}{2R}}_{\text{Tabor}}, \qquad R = 4.37 \mu m \qquad 2$$

This relationship assumes perfect sphericity of the indenter and a linear elastic relationship between the depth and diameter, which is not necessarily true. An alternative could be to use Tabor's [24] equation, which considers the elastoplastic response, or directly measure the depth using atomic force microscopy (AFM). The Spherical Tank relationship was used to calculate the crater volume and dynamic hardness values in Table 1. This is only an approximation because, at the rest depth, the pendulum loses contact with the indenter and thus cannot be accurately identified from the rebound or the second impact's velocity due to the 'ringing' from the dynamic compliance of the pendulum, as discussed earlier. The volume was approximated assuming a spherical shape of the crater [25], where:

$$V_{\text{crater}} = \frac{1}{6}\pi h \left(\frac{3}{4}D^2 + h^2\right) \qquad 3$$

Table 1: Approximated dynamic hardness ($H_D$) for different acceleration loads used during impact testing.

| Acceleration load (mN) | $D$ (μm) | $v_{in}$ (μm/s) | $v_{out}$ (μm/s) | $H_D$ (GPa) |
|---|---|---|---|---|
| 5 | 3.33 ± 0.13 | 1109 ± 1 | 828 ± 6 | 44.7 ± 1.4 |
| 8 | 3.88 ± 0.15 | 1369 ± 1 | 971 ± 2 | 40.9 ± 0.7 |
| 13 | 4.38 ± 0.16 | 1696 ± 2 | 1143 ± 5 | 41.3 ± 0.8 |
| 21 | 5.00 ± 0.15 | 2073 ± 2 | 1314 ± 12 | 38.7 ± 0.9 |
| 34 | 5.65 ± 0.15 | 2591 ± 2 | 1360 ± 5 | 42.9 ± 0.5 |
| 55 | 6.64 ± 0.13 | 3086 ± 2 | 1576 ± 4 | 29.9 ± 0.3 |
| 100 | 7.89 ± 0.12 | 3873 ± 6 | 1980 ± 9 | 19.3 ± 0.6 |

As shown in Table 1, a higher acceleration load increased the crater volume. The approximated dynamic hardness decreased with the acceleration load and was generally higher than the static hardness of 27 ± 3 GPa measured using conventional nanoindentation. This is due to the substrate contribution, as confirmed by FIB-SEM cross-sectioning, which increased as the crater depth increased.

The ability of the coating system to dissipate the impact energy can be calculated as the percentage of the dynamic energy change/loss to the input dynamic energy before the indenter reaches the surface, that is, the maximum dynamic energy at the maximum velocity. This relationship reflects the degree of plastic deformation during the impact. This means that the dissipated energy in the system (i.e., the composite of the coating and substrate) during impact can be expressed as follows:

$$E_{I,dis}(\%) = \left[1 - \left(\frac{v_{out}}{v_{in}}\right)^2\right] * 100 \qquad 4$$

where $v_{in}$ and $v_{out}$ are the velocities at which the indenter approaches and recedes from the surface after impact, respectively, calculated from the change in depth with time [15].

As shown in Figure 3, the percentage of dynamic energy change/loss increases with the accelerating load, with approximately 40% of the energy dissipated at 5 mN, which leaves an impact with a depth of approximately 500 nm. At low loads, the response was mainly due to the coating, as verified from FIB cross-sectioning, and at higher loads, the influence of the substrate became more prevalent. Overall, the amount of plastic work increased with the impact velocity, which is consistent with the strain-rate dependency of the hardness of the coating-substrate system. For metals, this was linked to the possibility of triggering a higher density of dislocations, resulting in hardening [26].

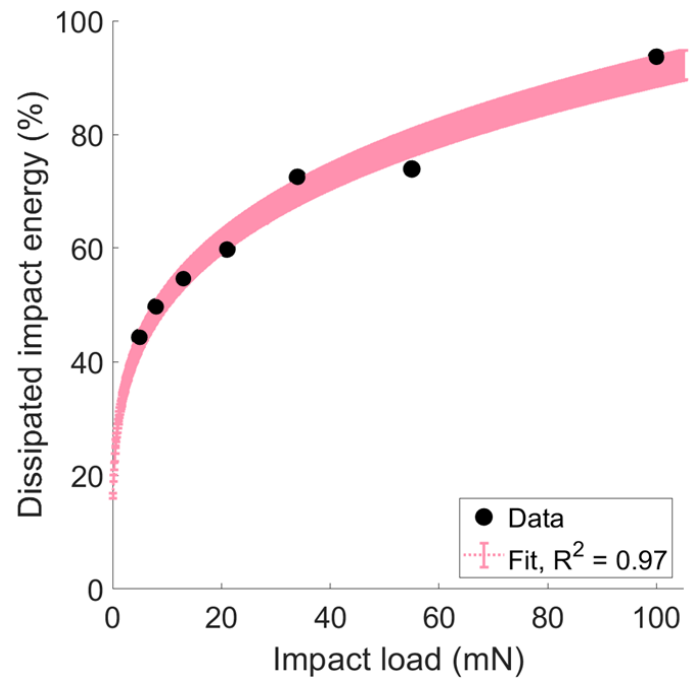

Figure 3: Energy dissipated ($E_{I,dis}$) from dynamic impact testing as a function of impact acceleration load ($L$). The fit is $E_{I,dis}(\%) = 29L^{0.25}$.

## 3.2. Cyclic impact test

The diameter of the impact crater increased with increasing number of cycles, indicating that the coating-substrate system underwent permanent plastic deformation during each cycle (Figure 4a). This plastic deformation accumulated as cycles increased, leading to coating failure at the point of impact (Figure 5a), accompanied by coating chipping radially around the impact crater (Figure 4b). In addition, the magnitude of the increment in the diameter of the impact crater depended on the load. This indicates that the ability of the coating-substrate system to withstand cyclic loading can be load-dependent at higher loads, as it results in more significant damage to the coating-substrate system from the first impact. Thus, the relationship among the number of repeated impacts (cycles), acceleration load, and impact crater diameter can describe the mechanical behaviour of the coating-substrate system under cyclic loading.

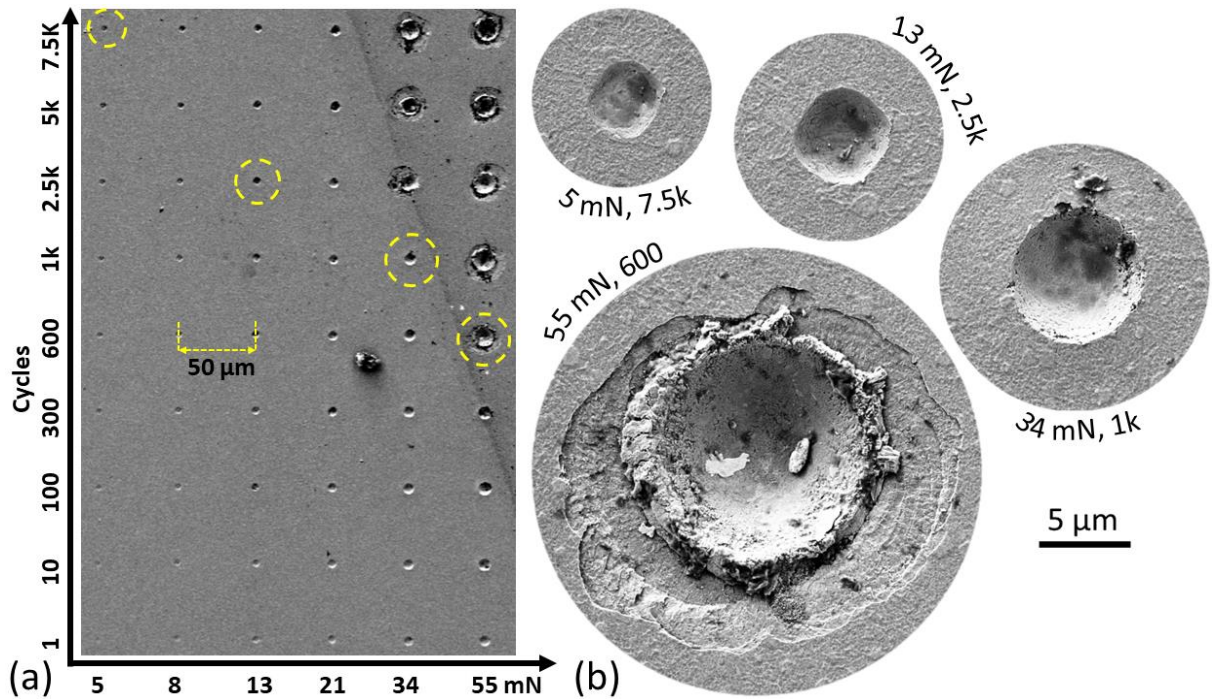

Figure 4: Low-voltage secondary electron image (SEM) of the craters of repeated impact tests on TiN coating with a diamond spherical indent for a range of cycles and acceleration load at room temperature. (b) SEM images for the impact sites' microstructures with the load (mN) and number of cycles indicated on the image correspond to locations circled in (a).

Selected impact craters were cross-sectioned, and chemical analysis for the cross-section was performed using energy-dispersive X-ray spectroscopy (EDX) at 10 kV, which showed a 200 ± 50 nm thick interlayer (interphase) between the coating and substrate. The interphase was found to be a titanium-iron compound, and the concentration of titanium increased gradually in the direction of the coating, and the iron concentration decreased, which was from when pure Ti was deposited before initially. The interphase also contained a slight nitride trace (Figure 5b-d).

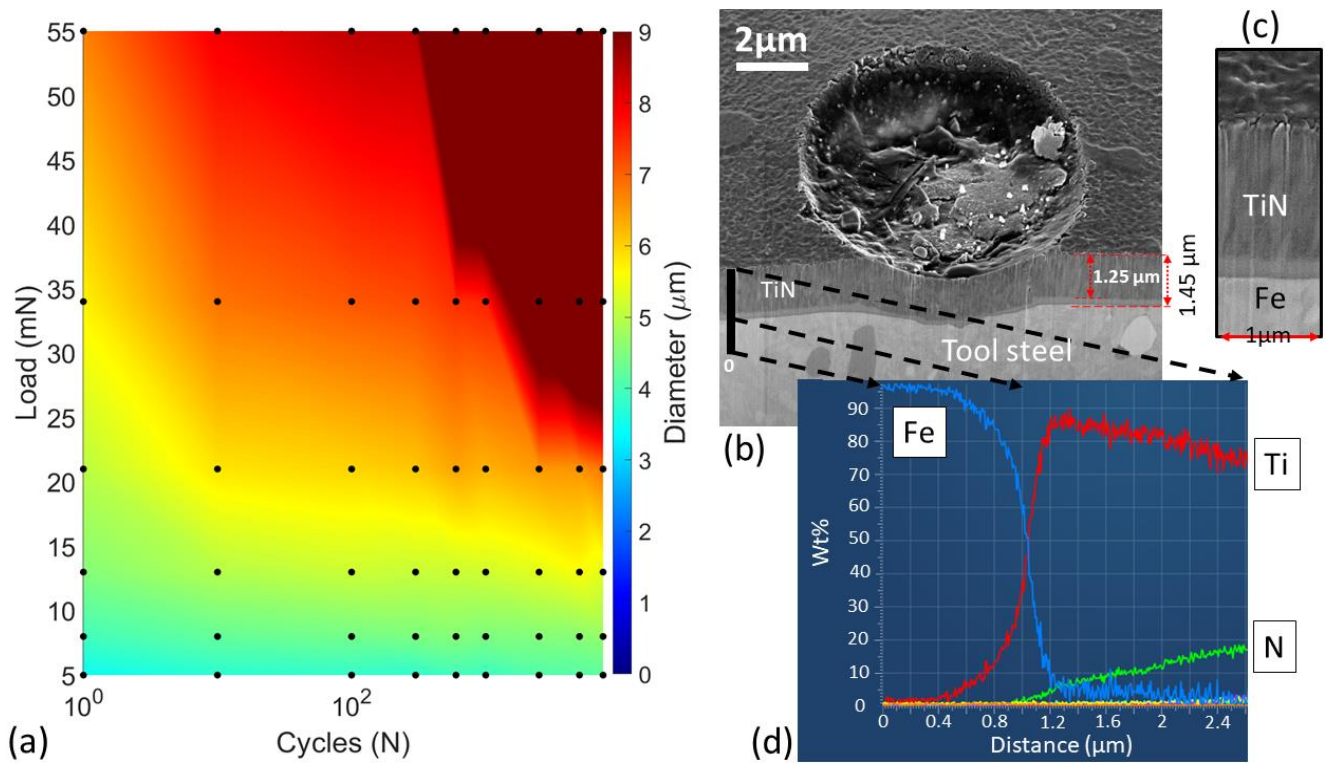

Figure 5: (a) Impact crater's diameter on the TiN surface after cyclic impact testing. The diameter was measured from high-resolution scanning electron images using ImageJ. (b) SEM image of the cross-section. (c) Higher-resolution SEM image showing a 200 ± 50 nm interphasial layer between TiN and the tool steel substrate. (d) SEM Energy Dispersive X-Ray (EDX) line profile through the cross-section shows the thickness of the interphase between the substrate (tool steel) and coating (TiN). The approximately 200 ± 50 nm thick interphase is a titanium and iron compound gradient with a slight nitride trace. Oxford Instruments and Aztec software were used for the EDX analysis.

Cross-sections of the impact sites showed the degree of damage to the coatings as a function of cycles and acceleration load (Figure 6). However, the extent of plastic deformation in the substrate could not be assessed using electron backscatter diffraction because of the small grain size of the tool steel, which affects the estimation of geometrically necessary dislocations [27]. The images also indicate that the impact-induced damage led to localised radial pile-ups around the impact crater (Figure 6a) and radial cracks before the coating chipping and extrusion of the substrate, eventually leading to coating failure from repetitive impacts (Figure 6b).

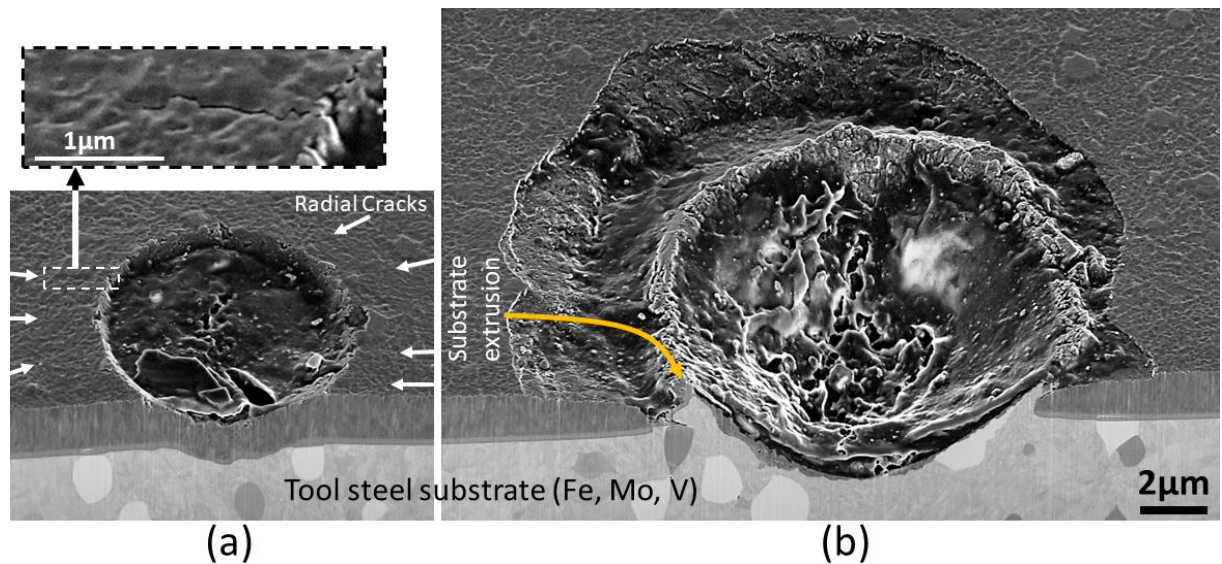

Figure 6: FIB cross-sections of repetitive impact sites. Impact craters using (a) 34 mN for 1000 cycles. (b) 55 mN for 2500 cycles. Both (a) and (b) are on the same scale. Arrows indicate the sites of radial cracks, magnified at the top of (a).

A three-dimensional map was generated from the acceleration load, sensed depth, and number of cycles, including at which the coating failed, as shown in Figure 7a. Coating failure or 'point-of-failure' was defined as an abrupt change in measured depth during the test, such as a jump in the measured depth from approximately 1.4 µm to 2 µm (Figure 7), later confirmed by cross-sectioning and SEM imaging. The displacement was recorded with time (cycle at 4 s intervals) during repeated impact tests, which allowed the impact test depth to be measured as a function of cycles for each acceleration load. As shown in Figure 7, the depth increased incrementally with impact cycles. By contrast, the depth increased more significantly with the acceleration load when the number of cycles was constant.

The point of failure for different acceleration loads was around 2 µm depth, larger than the coating thickness of 1.5 ± 0.5 µm, indicating simultaneous deformation of the coating and the substrate. The plotted data shows three distinct response zones: blue at a depth of less than 0.5 µm, green at a depth of less than 1 µm, and amber at a depth of less than 1.5 µm. From post-mortem imaging and cross-sectioning, the blue zone corresponds to no interaction with the substrate, with the deformation being locally sustained by the coating. Interaction with the substrate starts in the green area of Figure 7b, as shown in Figure 6a, for 1000 impacts using an acceleration load of 34 mN. The yellow-reddish area corresponds to the coating failure, as shown in Figure 6b.

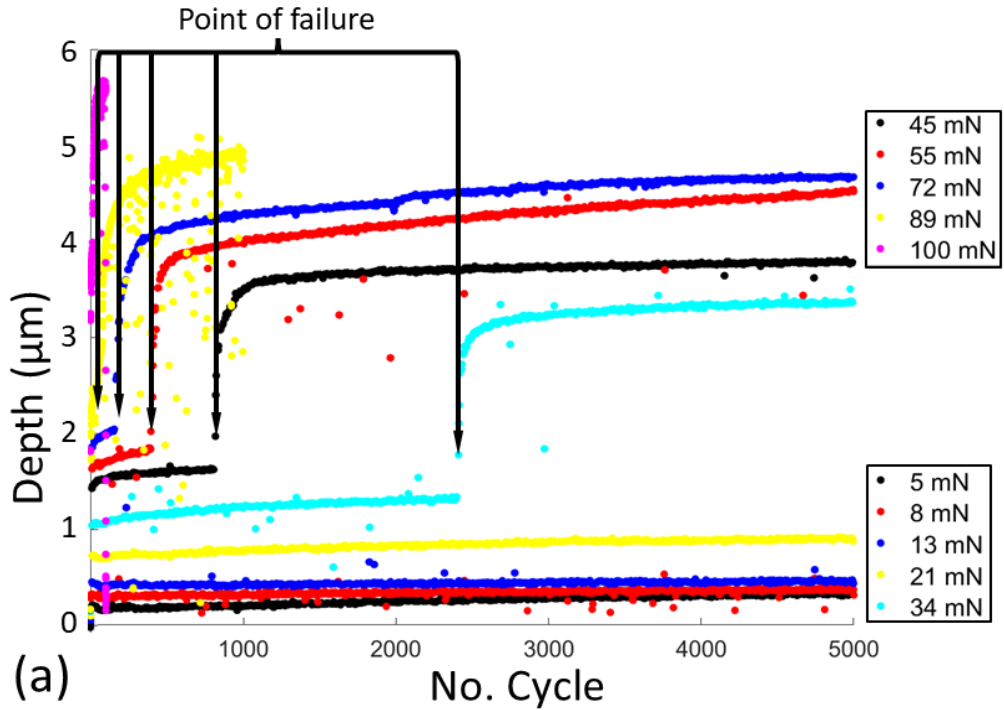

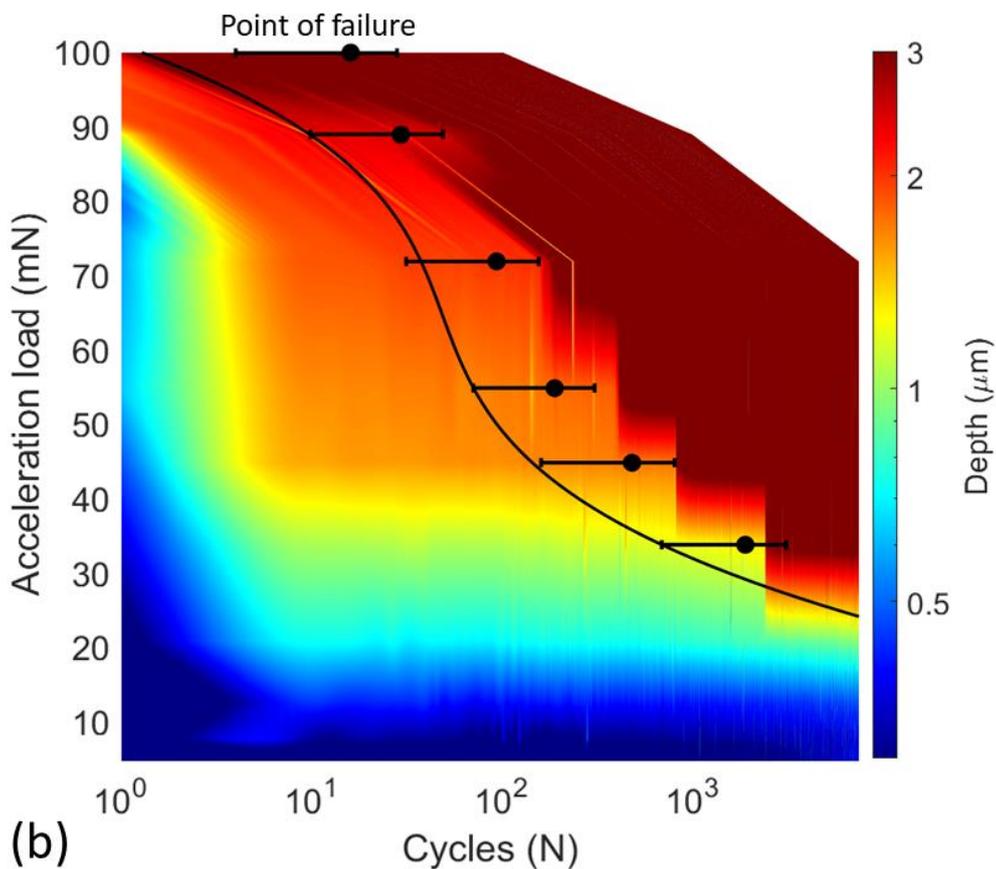

Figure 7: Impact depth data from multi-cycle micro-impact testing on TiN coating using a range of acceleration loads and up to 7500 cycles; (a) raw experimental data and (b) data interpolated into a three-dimensional map and fitted with equation 6. The test for a lower acceleration load was repeated (Figure 3) and consolidated. Depth or impact depth relates to a portion of the total length that the indent travels in and out of the sample, defined by the point at which the software decides where the zero is, which is not necessarily the original surface [23]. An abrupt depth jump during the test defined the 'point of failure'.

The resulting impact craters exceeded the size of the indenter (4.37 µm radius as shown in Figure 5a, and 1.8 µm depth as shown in Figure 7) due to damage accumulated (Figure 4b at 5mN and after 7,500 cycles). This is because as the impact progresses, the shape of the indenter crater does not remain spherical due to the high stresses involved (see Figure 4b at 55 mN and after 600 cycles), as high stresses accumulating from cycles lead to larger impact craters and more synergic response from the coating system, i.e., the TiN coating and steel substrate. However, the initial shape and size of the indenter are critical for the start of the impact process and the subsequent deformation behaviour, as different tests may produce data with different fundamental dimensions [28], [29] and failure mechanisms [30], [31].

### 3.3. Repetitive impact analysis

From the recorded depth during repeated impacts, the relationship between the depth ($h$) and cycles (N) before failure can be fitted to an empirical equation, as shown below:

$$h(N) = a\sqrt{N} + c \qquad 5$$

where c describes the depth at the first impact, and $a$ represents the incremental depth gradient $dh/dN$, during the cyclic impact. Both parameters are also functions of the acceleration load. The fitting was limited to data before the coating failure, defined by the abrupt increase in the depth measurement; for example, for the 34 mN test, that data was fitted to the first 2378 impacts.

The relationship between the acceleration load and depth at the first impact showed a positive correlation, which was nearly linear ($R^2$ = 0.95, Figure 8a). The values of these fitting constants describe the coating-substrate system's combined resistance, that is, dynamic hardness and toughness, with higher loads causing deeper craters and lower loads causing shallower craters. The scatter in the data is due to the sensitivity of the depth measurement of the initial distance between the indenter and the sample surface, as discussed earlier.

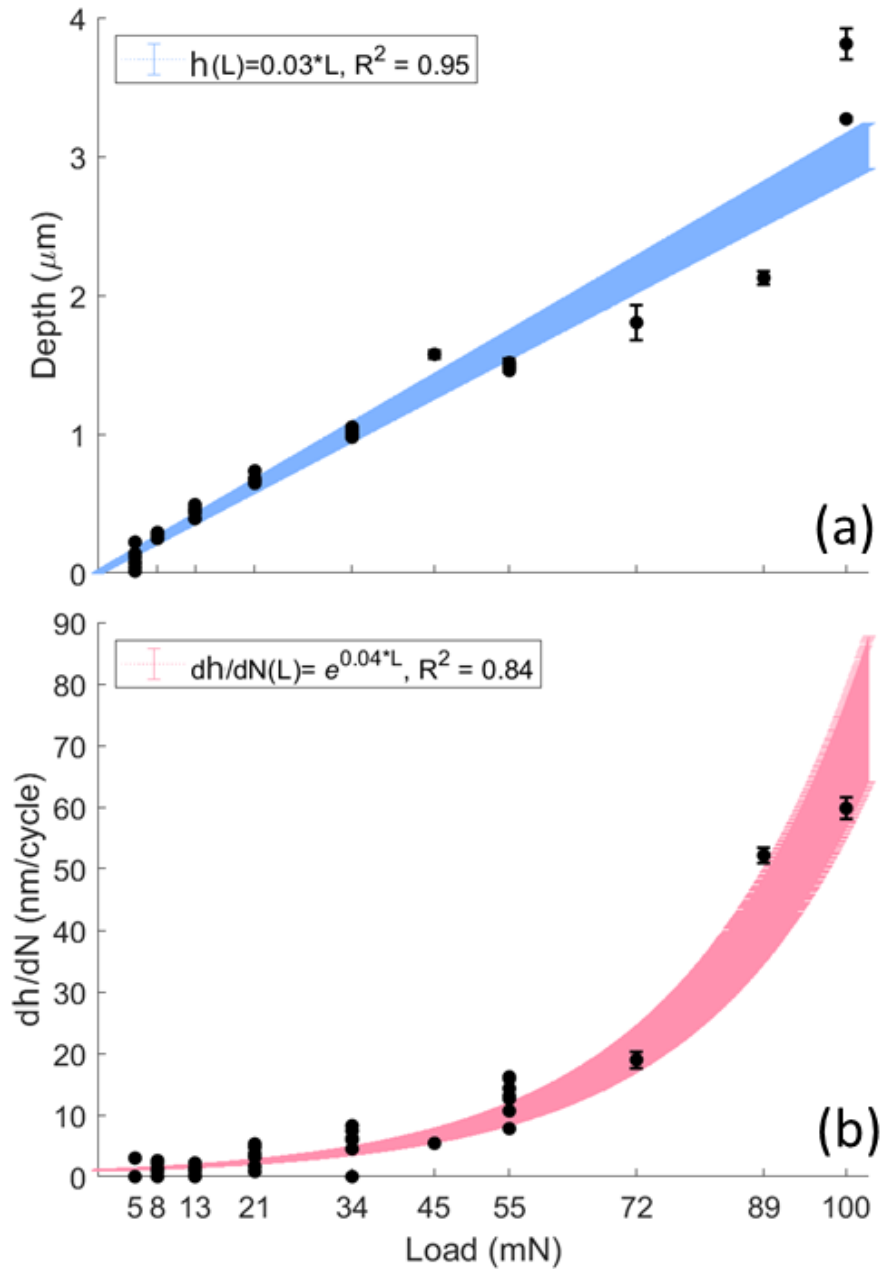

Figure 8: Fitting data from repetitive impacts on TiN coating as a function of spherical indenter acceleration load and indenter depth during the test. The experimental data and fitting for (a) the crater depth from the first impact and (b) for the depth increment or gradient ($dh/dN$) from repetitive impacts.

The relationship between the acceleration load and depth gradient, $dh/dN$, as shown in Figure 8b, can be used to understand the mechanical properties of the coating-substrate system after the first impact as a measure of damage accumulation before failure [32]. A graph plotting the acceleration load against $a$ shows a non-linear relationship between the two. The intercept on the y-axis at a load equal to zero can be interpreted as the residual plasticity, which can provide information about the initial residual stress of the coating-substrate system, especially at low loads, where the gradient is highly sensitive to coating initial residual stress and topographies.

By empirically fitting the experimental data, the relationship that governs the test, including the damage after the first impact and the accumulation of damage, can be described as follows:

$$h(L, N) = e^{0.04L}\sqrt{N} + 30L \qquad 6$$

Equation 6 (Figure 7b) includes two aspects of fatigue damage: the impact load, which determines the transition from a membrane-like failure of relatively thin coatings to plate-like failure in relatively thick coatings [33], [34], and the accumulation of damage during fatigue testing. Similar relationships between the load and depth of an indentation caused by a spherical indenter can be described mathematically using the mechanical properties of a material, such as its elastic modulus, yield strength, and plastic flow. A standard model used to describe this relationship is the Oliver-Pharr model [35]; however, this model simplifies the complex mechanical behaviour of materials under indentation and does not consider damage accumulation during repeated impact testing.

## 3.4. Computational model

It is important to investigate the distribution of the traction components along the interface during loading and unloading. Figure 9a shows the distribution of the normal ($T_n$) and tangential traction ($T_t$) components along the $r$-direction (see Figure 1b) for $L = 100$ mN. Traction is defined by $T_i = \sigma_{ij} n_j$ where $\sigma_{ij}$ is the stress tensor, and $n_j$ is the normal vector. Thus, for the interface between the film and the substrate, the unit normal is $n_z$ and the normal and tangential tractions are $T_n = \sigma_z$ and $T_t = \sigma_{rz}$, respectively.

The results show that the normal traction is compressive during loading, leaving residual tensile stress after unloading (Figure 9a). Hence, the tensile component contributes to the failure of the interface with repeated impacts. It should be noted that the representative amplitude of fatigue should be equal to the traction value after unloading because compressive traction does not directly contribute to the degradation of the interface [36], [37]. A similar behaviour was observed in the case of shear traction, as it changed its direction between loading and unloading, where the maximum magnitude occurred at the loading peak. Thus, the degradation was associated with the maximum magnitude. The magnitude of the shear traction is smaller than that of the normal component, which suggests that normal traction is responsible for the degradation.

Figure 9b and c illustrate the distribution of the normal traction after unloading and shear traction at the peak of the loading, which is the loading responsible for coating degradation. The results showed that the traction increased with the indentation load. The position of the maximum magnitude did not change significantly with an increase in the load. It should be mentioned that the maximum values were used to examine the correlation with fatigue life. Under such circumstances, the maximum stress extends deep into the substrate. The results may be affected more significantly by the hardness and toughness of the substrate, a phenomenon that is also observed in erosion tests under harsh conditions [38].

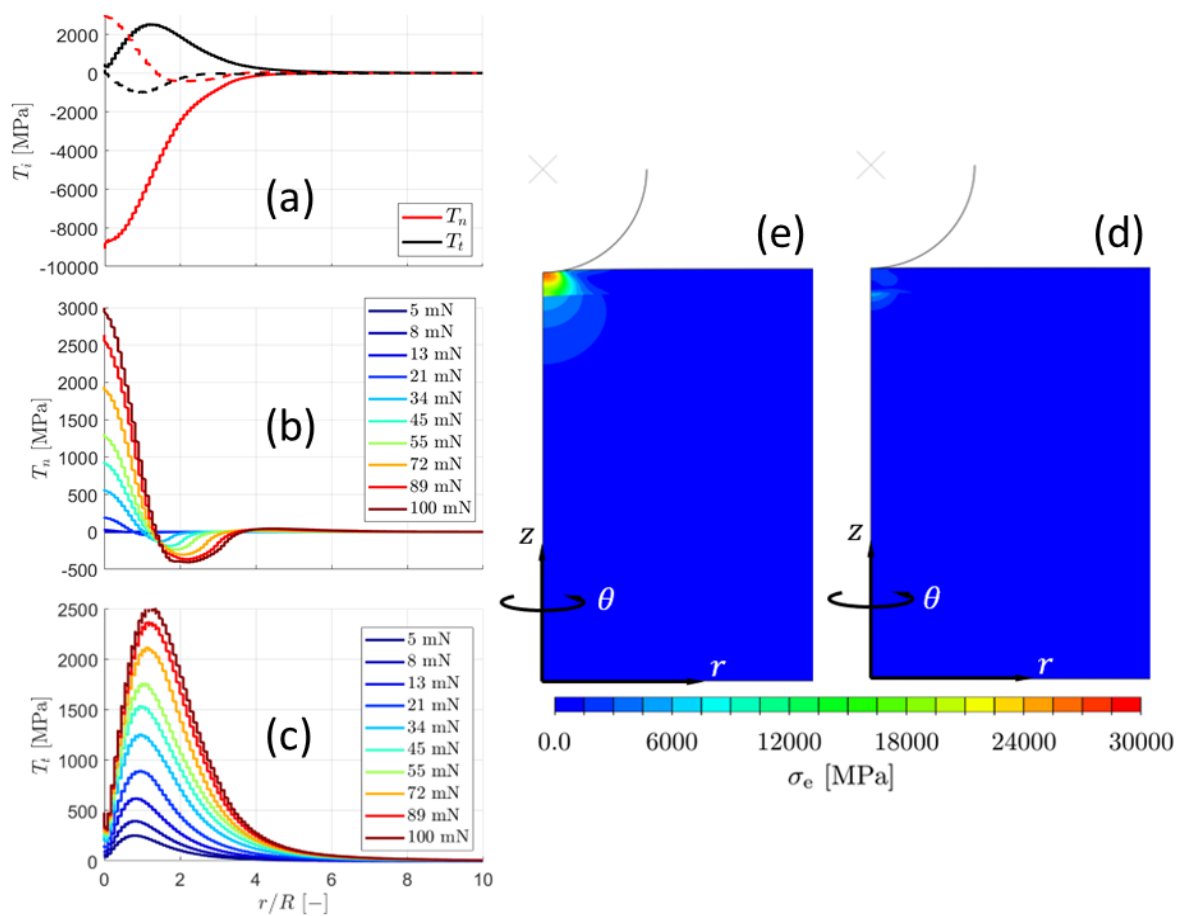

Figure 9: (a) The distribution of the normal and shear traction components, $T_n$ and $T_t$ along the interface at the peak of the load and after unloading, i.e., full and dashed lines, respectively, in the case of $L = 100$ mN. The distribution of the traction components along $r$-direction at different indentation loads: (b) the normal traction, $T_n$, after unloading; and (c) the shear traction, $T_t$, at the peak loading. Distribution of the equivalent von Mises stresses for L=100 mN and (e) at the peak of the load and (d) after unloading.

The deformation induced on the coating can be expressed in terms of equivalent von Mises stress. Figure 9e and d show the distribution of the equivalent von Mises stress at the coating-substrate interface at the maximum loading point in the case of $L = 100$ mN and at the peak of loading and after unloading. The distributions imply that the maximum values occurred at peak loading, and the residual stresses yielded relatively small values. The maximum value occurs under the indenter (i.e., $r = 0$).

## 4. General Discussion

A coating/substrate system typically undergoes complex loading conditions, where hard coatings used as a protective layer for the substrate are more easily damaged during applications under high loads, frequent loading, and high temperatures. For instance, the tensile residual stress from deposition or accumulation during the coating operation is often detrimental. When the elastic limit is exceeded, cracking occurs on the surface of brittle coatings or at the coating/substrate interface, sometimes described as an adhesion failure. Nonetheless, thin coating failure in cyclic micro-impact testing could be defined simply by an abrupt jump in depth-sensing during the test, as shown in Figure 7. According to the literature, this event consistently occurs in other coatings but is less pronounced in ductile coatings [11], [16], [23], [39], [40], [41], [42]. This definition of coating failure does not address the failure mechanism, which needs to be examined further. However, the definition provides a simple cut-off criterion, simplifying the construction of comparable S-N curves for coatings.

Therefore, the energy required for this *failure* event is a mixture of the cohesive failure of the coating system (i.e., the composite of the coating and substrate) through fracture and adhesive failure between the coating and the substrate, which depends on many factors, including the test setup. For example, the current setup, that is, perpendicular impact testing, emphasises the role of the coating fracture toughness, as this sudden jump in depth or *failure* can be mainly due to coating cracking or chipping (i.e., cohesive failure) as well as adhesion toughness [43]. Thus, this testing setup is unsuitable for evaluating the adhesion of brittle coatings on ductile substrates because the brittle coating cracks before delamination occurs, and it is not ideal for testing strongly adhered coatings for the same reason [44], [45].

A potential way of improving the test is to use an inclined setup for cyclic impact testing. This would ensure the inducement of more shear forces within the coating and interface, which makes the test more appropriate for gauging the practical adhesion strength, as coating failure in this scenario will be an assured synergy between the interphase toughness and coating toughness [36], [46].

Although TiN is known to have excellent adhesion, its fracture toughness is not very high relative to other materials [47], [48], [49]. Thus, considering the lower energy required for fracture compared to delamination, it was expected that coating failure would be cohesive, occurring because of coating cracking rather than delamination [49], as the current test setup emphasises compression (during loading) and tension (during unloading), but not delamination [50]. Nevertheless, as shown in Figure 6b, TiN coating failure was also related to the malleability of the steel substrate, as it was extruded (evidenced by the pile-up) during the cyclic impact testing (Figure 4a).

The findings of this study show that the fracture and cracking processes are caused by localised stress from surface impact. From the experiment and simulation, it can be inferred that repeated impacts initiated surface cracks. In contrast, forces imposed through the impact introduced a residual stress field that, in turn, caused the interface to delaminate. This led to a progressive reduction of the dynamic hardness due to the weakening of the initial coating-substrate interfacial strength and compressive stress severely localised deformation at the interface. Finally, when the combined critical limit of the maximum normal impact load and impact cycles is exceeded, the coating removal becomes dominant, which, when combined with substrate extrusion, leads to coating chipping and thinning by impact (Figure 10). Nevertheless, whether the thinning of the coating is due to localised plasticity or chipping, this thinning makes the TiN coating behave as a membrane with concentrated tensile stresses at the surface centre and interface centre, as reported in [33], [51]. The tensile stresses at the edges of the impact crater led to edge cracking, and the limited delamination that occurred owing to substrate extrusion could be attributed to exceeding the critical shear failure stress.

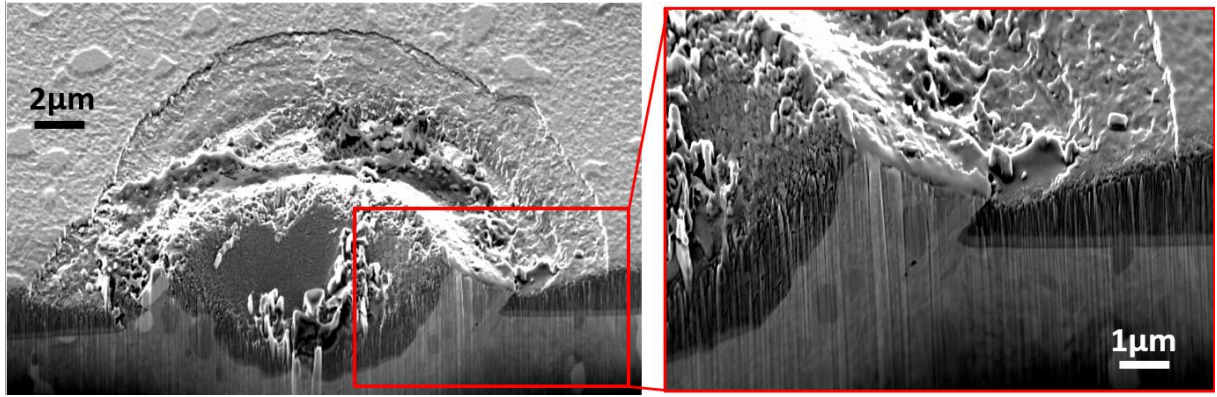

Figure 10: SEM image of a cross-section of the impact crater after 100 cycles using 100 mN acceleration load.

The current test can address the coating system (i.e., the composite of the coating and substrate) fracture toughness, as the fatigue failure modes of the thin coating are primarily due to coating chipping from the localisation of plastic deformation in the coating and minimal debonding of the interface between the coating and the substrate (Figure 6a). Interrogation of the data from further impact testing (Figure 7) using non-linear regression allowed for linking the parameters a and c to the load. The final relationship between the depth ($D$), load ($L$), and cycles ($N$) (i.e., $D(L,N)$), which describes the initial damage and accumulation of damage with cycles, can be fitted to an empirical equation, as shown in equation 6 and Figure 7b. However, caution must be exercised when interpreting the fitting parameters because the current fitting method does not address the physical description of the problem or the size dependency of the test.

Nonetheless, the current analysis has the potential implications for enhancing coating design and improving durability in practical applications, along with enabling more accurate prediction of the service life of coated components. In addition, future work should investigate the experimental separation of the loading and impact velocities, better depth sensing, and inclined impact testing. In addition, future work needs to delve into a mechanistic simulation of coating failure by debonding due to repetitive impacts.

# 5. Conclusion

This study examined the failure of a TiN coating on an ASP23 tool steel substrate under dynamic and cyclic loading conditions, revealing insights into the response and degradation of adhesion under impact and cyclic impact fatigue. This study highlighted the inverse relationship between the strain rate and the dynamic hardness for dynamic testing, with greater energy dissipation at higher acceleration loads.

Computational models provided insights into the distribution of traction components and stress at the coating-substrate interface, revealing a transition from compressive to tensile normal traction during unloading and emphasising the potential involvement of tensile stresses in fatigue-related degradation. The maximum equivalent von Mises stress was observed beneath the indenter during loading, indicating that critical regions were susceptible to material damage during cyclic loading.

In the cyclic impact tests, the TiN coating was subjected to repetitive micro-impacts across a range of acceleration loads, simulating various fatigue conditions. While the coating exhibited a linear response during the first impact, the accumulation of damage during repetitive impacts followed a non-linear pattern. This leads to the build-up of residual compressive stress and plasticity, ultimately resulting in coating failure identified by an abrupt change in the sensed impact's crater depth.

# Acknowledgements

The authors thank Andy Fox (Wallwork Group Ltd) for supplying the coating, Dr Ken Mingard (National Physical Laboratory) for proofreading the article, and the National Measurement System (NMS) programme of the UK government's Department for Science, Innovation and Technology (DSIT) for financial support.

# CRediT author statement

**Abdalrhaman Koko:** Conceptualisation, Methodology, Visualisation, Investigation, Formal analysis, Writing - original draft.

**Elsiddig Elmukashfi:** Investigation, Visualisation, Writing - original draft.

**Tony Fry:** Resources, Writing - review & editing.

**Mark Gee:** Writing - review & editing.

**Hannah Zhang:** Methodology, Writing - original draft.